\documentclass[mathleft]{an_a_astroph}
\usepackage{graphicx}
\usepackage{times}
\usepackage{natbib}

\newcommand{\teff}{\mbox{${T}_{\rm eff}$}}

\newcommand{\msol}{\mbox{${\rm M}_{\odot}$}}

\newcommand{\simgt}{\lower.5ex\hbox{$\; \buildrel > \over \sim \;$}}
\newcommand{\simlt}{\lower.5ex\hbox{$\; \buildrel < \over \sim \;$}}
\newcommand{\BV}{Brunt-V\"ais\"ala\ }

\begin{document}


\title{The effect of turbulent mixing on the g-mode spectrum of MS stars}

\author{J. Montalb\'an\fnmsep \thanks{\email{j.montalban@ulg.ac.be}}
\and A. Miglio
\and P. Eggenberger 
\and A. Noels
}
\titlerunning{The effect of turbulent mixing on the g-modes}
\authorrunning{Montalb\'an et al.}
\institute{Institut d'Astrophysique et G\'eophysique de l'Universit\'e de Li\`ege, All\'ee du six Ao\^ut, 17 B-4000 Li\`ege, Belgium}



\abstract{ The understanding of transport processes inside the stars is one of the main goals of asteroseismology. Chemical turbulent mixing can affect the internal distribution of $\mu$ near the energy generating core, having an effect on the evolutionary tracks similar to that of overshooting. This mixing leads to a smoother chemical composition profile near the edge of the convective core, which is reflected in the behavior of the buoyancy frequency and, therefore, in the frequencies of gravity modes. We describe the effects of convective overshooting and  turbulent mixing  on the frequencies of gravity modes in B-type main sequence stars. In particular, the cases of  p-g mixed modes in $\beta$~Cep stars and  high-order modes in SPBs are considered.}

\maketitle

\section{Introduction}

The standard stellar evolution modeling includes only the mixing-length theory \citep{Bohm58} to describe the convective mixing, and, for low mass stars, also the effect of microscopic diffusion. Contrarily to predictions of this  modeling, some chemical transport must also occur in the stellar radiative regions if we want to account for some observational facts.
In particular, the comparison between theoretical models and observational data from binaries and stellar clusters \citep[see e.g.][]{Andersen90,Ribas00} has shown that the  standard model underestimates the mixing in the central regions of the stars. It is largely  accepted that some extra-mixing takes place just at the border of the convective core that develops for stellar models with masses larger than $\sim~1.2$~\msol, but there is not consensus about the physical processes responsible for this mixing: convective overshooting \citep[e.g.][]{Schaller92}, microscopic diffusion \citep{michaud04}, rotationally induced mixing \citep[see e.g.][and references therein]{MM00,Maeder03}, mixing by non-adiabatic propagation of gravity waves excited at the boundary of the convective core \citep{Young05}, etc.

Generally, stellar modeling includes an overshooting parameter ($\alpha_{\rm OV}$) to account for the extra-mixing above the limit of the convective core, whose value has been estimated by fitting of stellar clusters or binary systems.  Recently,  asteroseismic determinations of the overshooting parameter have also  been done by using the advantageous properties of low order p and g  modes detected in some  $\beta$~Cep targets \citep[see e.g.][]{Pamy04,Aerts03,Briquet07a}. In fact, the frequency of modes close to an avoided crossing \citep{avoidedcrossing} are very sensitive to the extension of the mixed region \citep[e.g.][]{Dziembowski91,aud0}. The shape of the composition transition zone is also a matter of great importance as far as asteroseismology is concerned. In particular it significantly affects the term $\nabla_\mu$ appearing in the \BV frequency and plays a critical role in the phenomenon of mode trapping. What we show in this contribution is that \citep[as already suggested by][for $\delta$~Scuti stars]{goupiltalon02} the oscillation spectrum of B-type stars is also very sensitive to the slope of the chemical gradient in the central region of the star, and therefore, to the kind of transport processes that produced it.

The models with overshooting and turbulent mixing we computed to analyze how different  extra-mixing processes affect the  oscillation frequencies are described in section~\ref{sec:stellarmodels}, and the effects of these transport processes on the stellar structure and on the spectrum are reported in section~\ref{sec:overtur}. In a first approach we consider a parametric turbulent mixing without caring about its physical origin. If this chemical transport were generated by rotation according to the formalism implemented in Geneva code \citep[see. e.g.][]{Eggenberger07}, for instance, the effects due to the angular momentum transport should also be taken into account in the oscillation spectrum. In section~\ref{sec:rot} we present an estimation of these effects  for a typical $\beta$~Cep star and in section~\ref{sec:conclu}  we give our conclusions.

\section {Stellar models}
\label{sec:stellarmodels}
To study the effects of different extra-mixing processes in SPB and  $\beta$ Cep   pulsators we focus on models of 6 and 10~\msol\ respectively.  As a first approach, the turbulent mixing has been modeled by a diffusion process with a parametric turbulent diffusion coefficient $D_{\rm T}$ that is constant inside the star and independent of age. This  is a very rough approximation without any a priori  physical justification. We computed also some models using the Geneva code that include the treatment of rotation and linked transport processes such as described in Maeder \& Zahn (1998) and Maeder (2003). The comparison between both series of models shows that chemical composition profiles in the central regions provided by a uniform and time independent diffusion coefficient represent a first approximation, at least for massive stars,  of the effect would be produced by such a rotationally induced  chemical transport.

\begin{figure}
\begin{center}
\includegraphics[width=8cm]{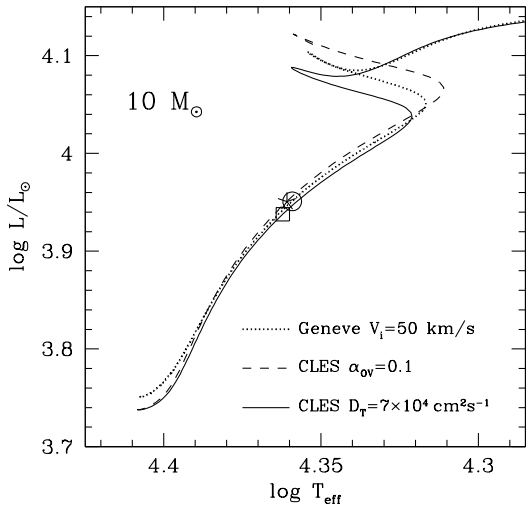}
\caption{ HR diagram showing main sequence evolutionary tracks of 10~\msol models computed with CLES evolution code with an overshooting parameter $\alpha_{\rm OV}=0.1$ (dashed line), and with a constant turbulent diffusion coefficient $D_{\rm T}=7\times10^{4}$ cm~s$^{-2}$ (solid line). The dotted line corresponds to the evolutionary track of a 10~\msol\ model computed with the Geneva evolution code assuming an initial rotational velocity $V_{\rm rot}=50$~km~s$^{-1}$. The symbols in the middle of MS track indicate the location of models with $X_{\rm C}=0.3$.}
\end{center}
\label{fig:hr}
\end{figure}

The models with overshooting were computed with CLES \citep[Code Li\'egoise d'Evolution Stellaire][]{ScuflaireCLES}. In this code the thickness of the overshooting layer $\Lambda_{\rm OV}$ is parameterized in terms of the local pressure scale height $H_p$: $\Lambda_{\rm OV}=\alpha_{\rm OV}\times(\min(r_{\rm cc},H_p(r_{\rm cc}))$, where $r_{\rm cc}$ is the radius of the convective core given by the Schwarzschild criterion and $\alpha_{\rm OV}$ is a free parameter. The values of  $\alpha_{\rm OV}$ used in this study were 0, 0.1 and 0.2.

\begin{figure}
\begin{center}
\includegraphics[width=8cm]{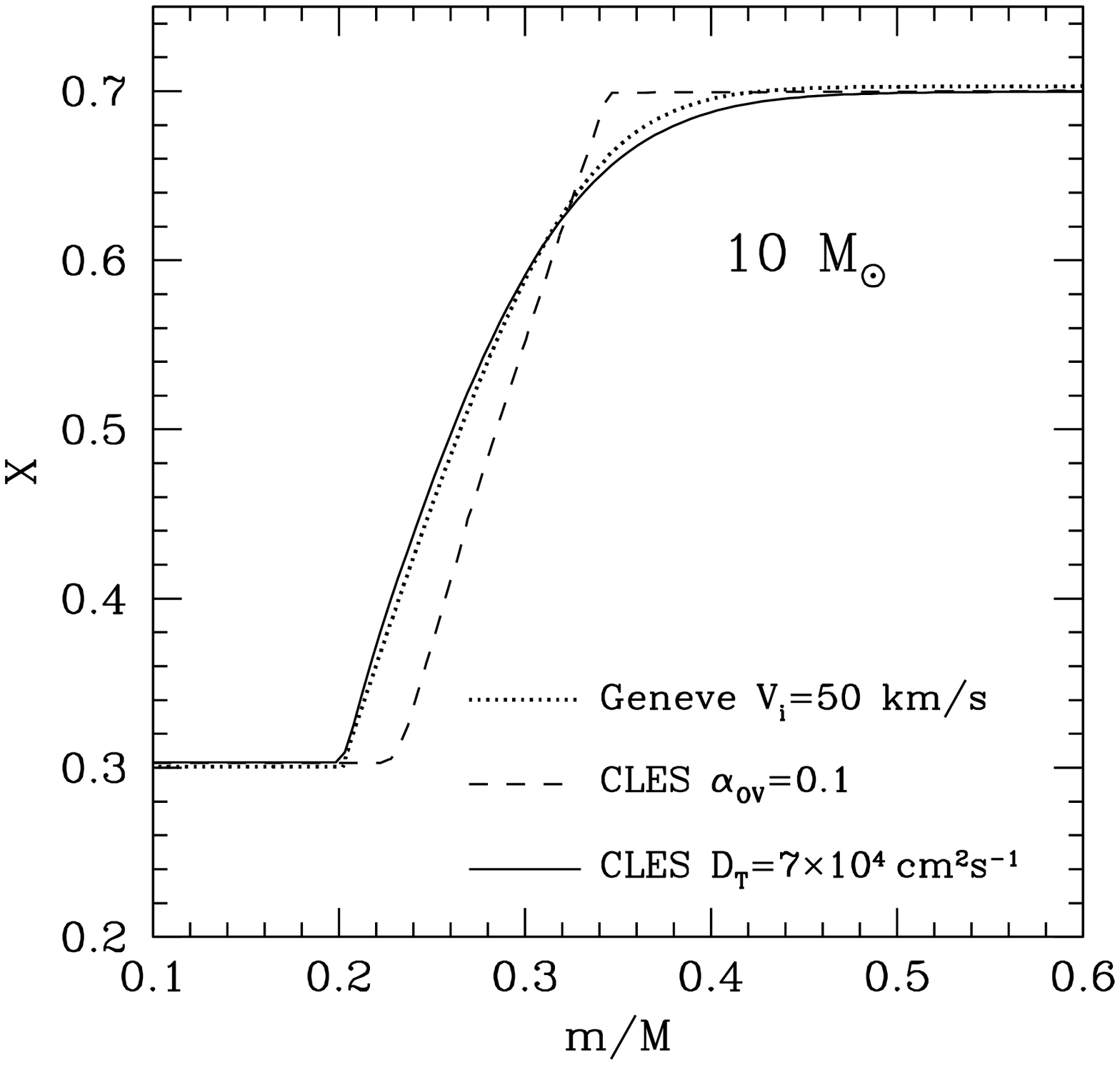}
\includegraphics[width=8cm]{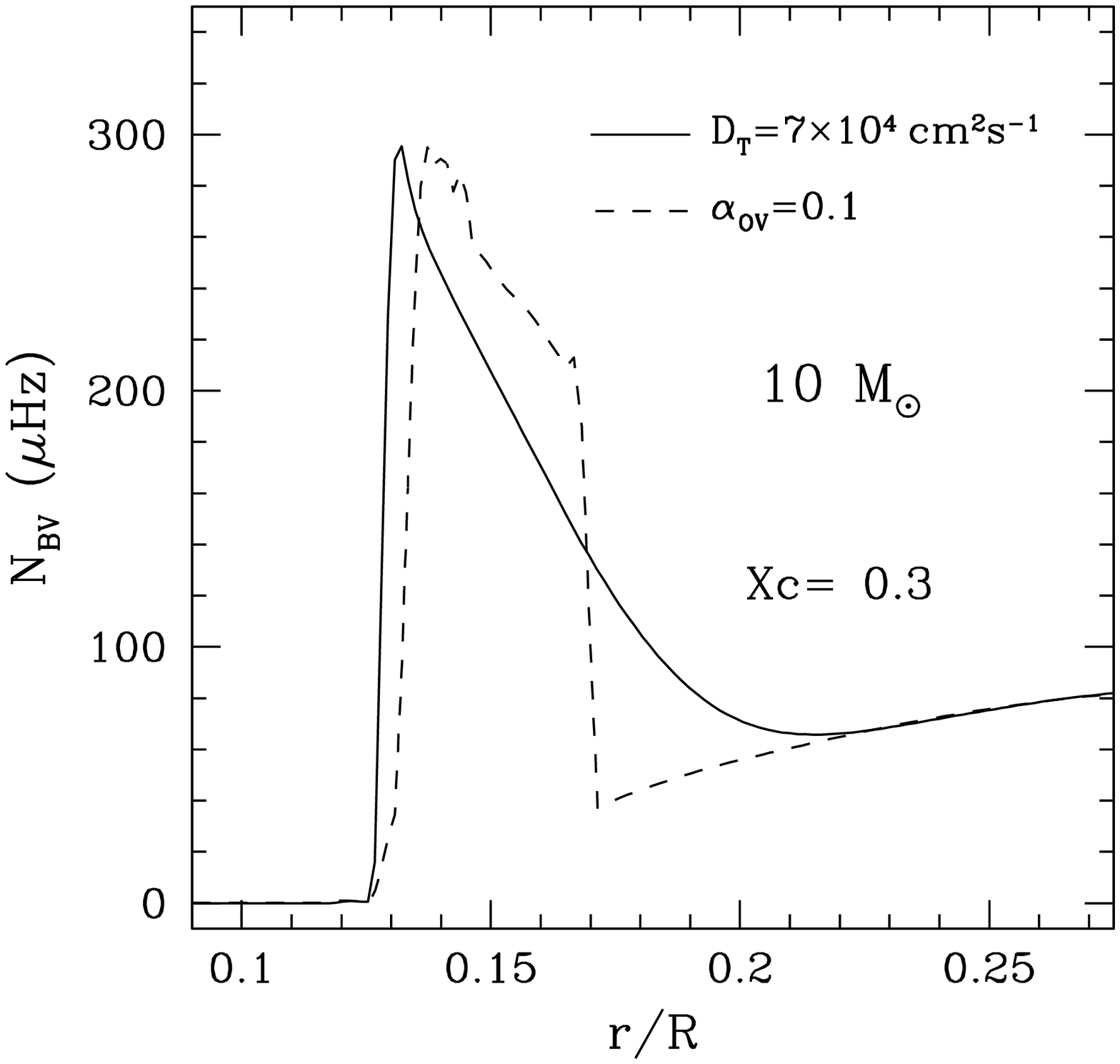}
\caption{Upper panel: Hydrogen abundance profile in the central regions of 10~\msol\ models with $X_{\rm C}\simeq0.3$. The different lines correspond to models computed with CLES, with overshooting (dashed line), and turbulent diffusion coefficient $D_{\rm T}$ (solid line), and with the Geneva code with an initial rotational velocity of 50~km~s$^{-1}$ (dotted line). Lower panel: \BV\ frequency profile in the central regions of the 10~\msol\ CLES models in upper panel.}
\end{center}
\label{fig:xprofil}
\end{figure}

The values of the parameter $D_{\rm T}$ were chosen in order to be close to the value of the chemical diffusion coefficient near the core  provided by the Geneva models. 
SPBs and $\beta$~Cep pulsators are considered as slow or moderate rotators. SPBs show a typical rotational velocity  of 25~km~s$^{-1}$ \citep{Briquet07}, whereas the range of projected rotational velocity in $\beta$~Cep stars extends from  0 to 300~km~s$^{-1}$ with an average of 100~km~s$^{-1}$ \citep{StankovHandler05}.   The Geneva  code calculations for 10~\msol\ models provide values of the chemical diffusion coefficient near the convective core with  $X_{\rm C}=0.3$,  of the order of   $5\times10^4$~cm$^2$s$^{-1}$  for an initial rotational velocity ($V_i$) of 20~km~s$^{-1}$, $7\times10^4$~cm$^2$s$^{-1}$ for $V_{i}$=50~km~s$^{-1}$, and $1.6\times10^5$~cm$^2$s$^{-1}$ for $V_i$=100~km~s$^{-1}$. On the other hand, the effect of an initial rotational velocity of 25km~s$^{-1}$ on the central hydrogen distribution of  a 6~\msol\ model is well mimicked by a $D_T \sim 5000$~cm$^2$s$^{-1}$. The results presented in this paper concern mainly the parametric models with $D_{\rm T}=5\times10^3$~cm$^2$s$^{-1}$ for SPB model and $D_{\rm T}=7\times10^4$~cm$^2$s$^{-1}$ for  $\beta$~Cep one, and they will be compared with overshooting models closely located in the HR diagram, that means  $\alpha_{\rm OV}$=0 for the SPB case and 0.1 for the $\beta$~Cep one.

\section{Overshooting versus turbulent diffusion}
\label{sec:overtur}

\subsection{Effects on the structure}

\begin{figure}
\begin{center}
\includegraphics[width=8cm]{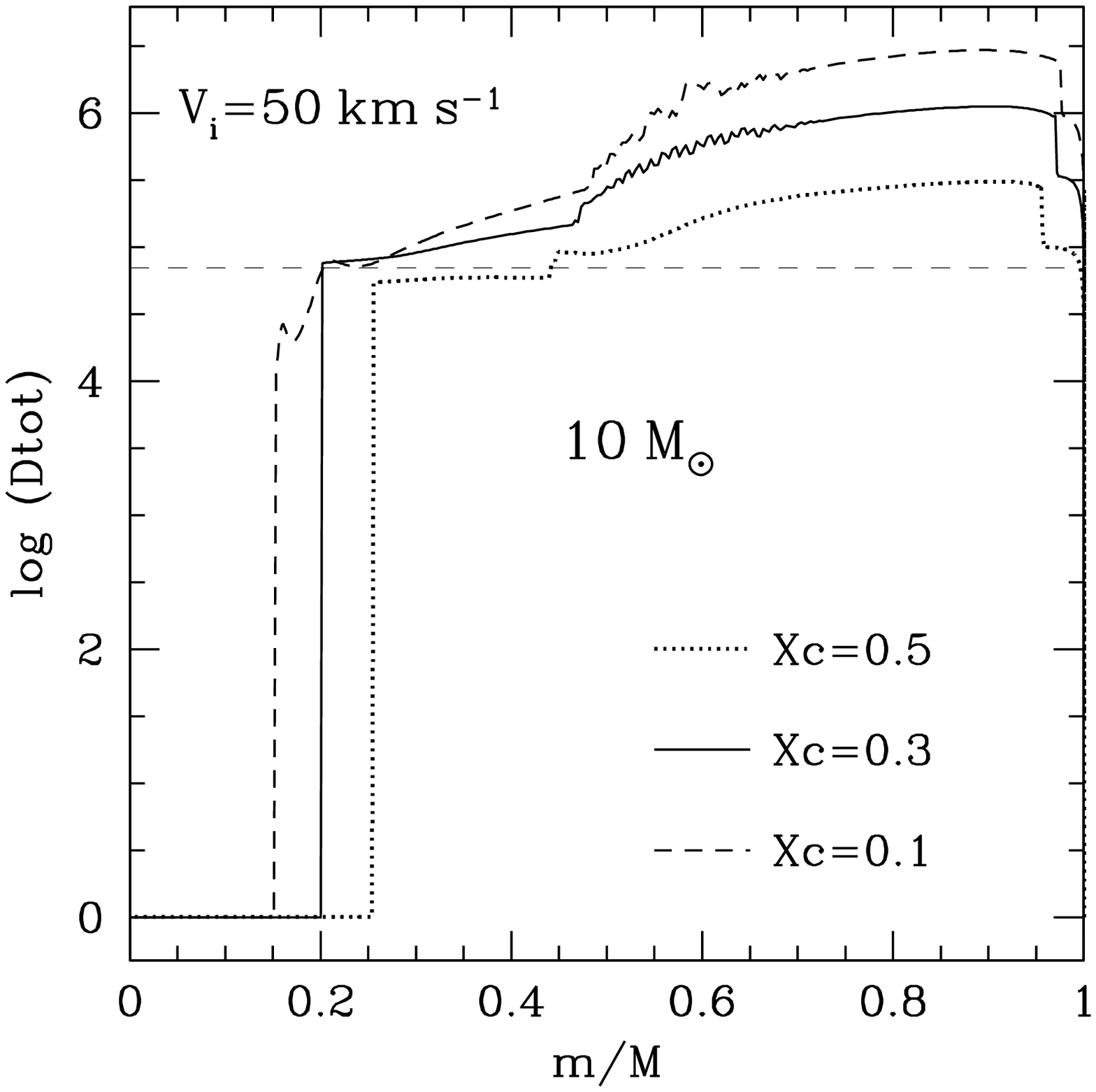}
\caption{Total chemical diffusion coefficient for a Geneva 10~\msol\ model at three different evolutionary stages, and with an initial rotational velocity $V_{\rm i}=50$~km~s$^{-1}$. The horizontal dashed line gives our chosen value for the computations with CLES.}
\end{center}
\label{fig:Deff}
\end{figure}

Whatever the origin of the chemical extra-mixing at the border of the convective core, the effect on the evolutionary track is an increase of luminosity and of the duration of the main-sequence phase. The evolutionary tracks of 10~\msol\ models with different treatments of chemical mixing are shown in Fig.~\ref{fig:hr}. Though the HRD location of  turbulent mixing models is quite close to that of models computed with overshooting, the chemical composition gradient near the core is significantly different (see Fig.~\ref{fig:xprofil} upper panel) and therefore also the properties of the \BV\ frequency (Fig.~\ref{fig:xprofil} lower panel). Note that the hydrogen abundance profile near the core that results from Geneva code (including a consistent  treatment of rotationally induced transport processes)  shows a behavior similar to that obtained by using the parameterized turbulent mixing in CLES. The values of the chemical diffusion coefficient induced by rotation in Geneva models for three different evolutionary stages are shown in Fig.~\ref{fig:Deff} for 10~\msol\ models with an initial rotational velocity $V_i=50$~km~s$^{-1}$. We note the slight temporal variation of  this diffusion coefficient near the convective core boundary ($m/M\sim0.15-0.25$) and the fact that the value of $D_{\rm T}$ we chose (horizontal line) rather corresponds to an average of Geneva model values.

\subsection{Effect on oscillation spectrum}

SPB and $\beta$~Cep stars show oscillation modes excited by the classical $\kappa$-mechanism in the  iron group opacity bump at $T \sim$200000~K \citep{Dziembowski93,Dziembowski93a}. While SPB stars pulsate with high-order g-modes with periods going from 1 to 3 days (frequency $\nu$: 4~--~12~$\mu$Hz), low-order g and p modes as well as g-p mixed modes with frequencies between 30 and 100 ~$\mu$Hz are found excited in $\beta$~Cep stars.
Miglio et al. (2007, these proceedings) show that,  even for slow rotation, the changes of molecular gradient induced by the turbulent mixing acting near the core lead to significant variations in the properties of  high-order g-modes spectrum.  Although the approximation used in Miglio et al (2007) to relate the sharpness of $N_{BV}$ with the properties of the g-mode spectra is no longer valid for low-order modes, the analytical description of gravity modes presented there is still able to qualitatively describe the properties of low-order g-mode spectra (see their Fig.~4).

As the star evolves, the combined action of nuclear reactions and convective mixing leads to a chemical composition gradient at the boundary of the convective core, an increase of the \BV\ frequency and therefore of the frequencies of gravity modes. The latter interact with pressure modes of similar frequency and affect the properties of non-radial oscillations by the so-called avoided crossing phenomenon. The  modes undergoing an avoided crossing (mixed modes) are therefore sensitive probes of the core structure of the star.

\begin{figure*}
\begin{center}
\includegraphics[width=0.70\textwidth]{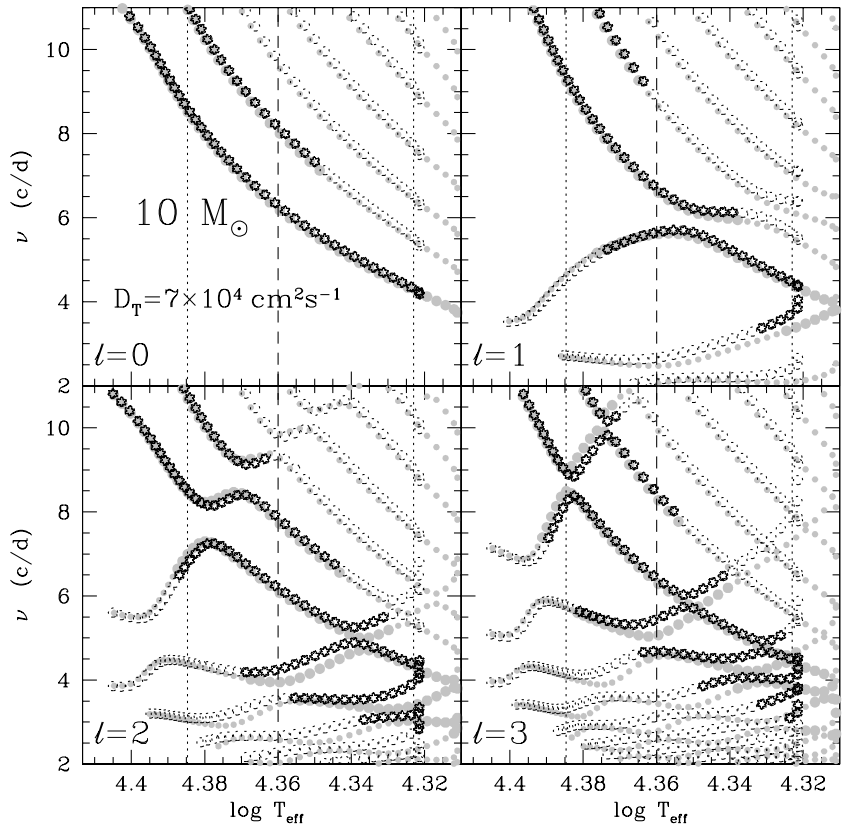}
\caption{Frequencies of  pulsation modes with angular degree $\ell=0-3$ as a function of $\log$\teff\ for main-sequence models of a 10~\msol\ star. Gray dots correspond to the frequencies of models computed with an overshooting parameter $\alpha_{\rm OV}=0.1$, whereas black circles are the frequencies of models computed with a turbulent diffusion coefficient $D_{\rm T}=7\times10^4$ cm~s$^{-2}$. Excited modes are represented by thicker symbols. The vertical lines indicate the effective temperature of models with an hydrogen  mass fraction at the center  of the order of  0.5, 0.3 and 0.1 (left to right).}
\label{fig:exciD7}
\end{center}
\end{figure*}

Here we study the effects of extra-mixing on the $\beta$~Cep spectra by comparing the properties of low-order g and p modes in models computed with overshooting and with chemical turbulent diffusion. Since the parameters $\alpha_{\rm OV}$ and $D_{\rm T}$ were chosen to lead to similar evolutionary tracks, such comparison allows us to remove the differences in the frequencies due to a different stellar radius. In Fig.~\ref{fig:exciD7} we plot the oscillation frequencies  for 10~\msol\ models along the main-sequence phase.
As expected, the differences  between frequencies of overshooting models and turbulent mixing ones is very small for pressure modes, while significant differences appear for mixed modes. These differences increase with the radial order and with the angular degree of the modes. In the figure we have also marked with thicker symbols the frequencies of  modes that the non-adiabatic oscillation code MAD \citep{Dupret03} predicts to  be excited. Here we show only the case with $D_{\rm T}=7\times10^4$~cm$^2$s$^{-1}$ but computations with lower or higher efficiency of the turbulent mixing show that the differences between overshooting and turbulent-mixing models increase with the value of $D_{\rm T}$.

\begin{figure}
\begin{center}
\includegraphics[width=8cm]{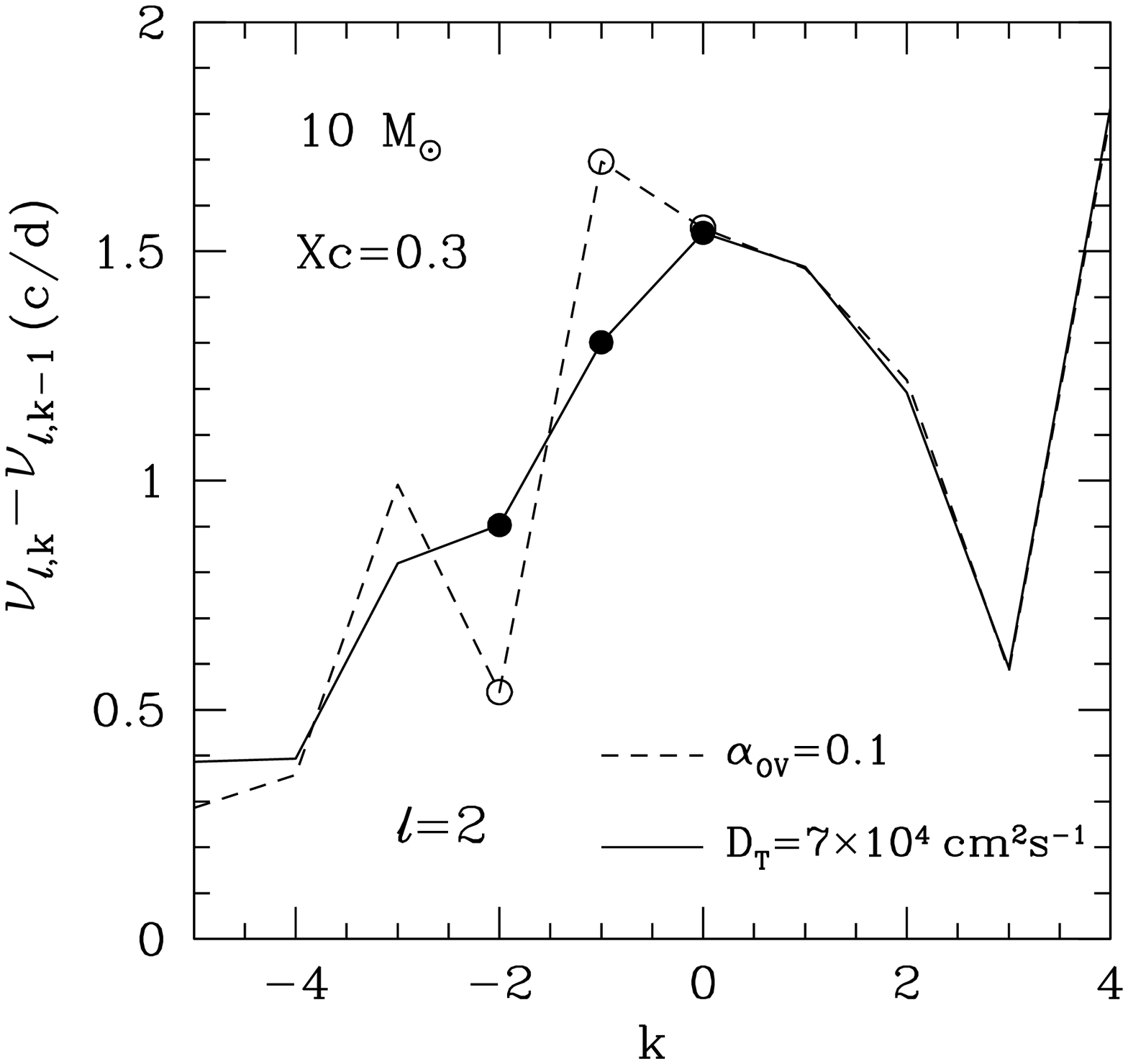}
\caption{Frequency difference (in c/d) between modes with consecutive radial order ($k$) and degree $\ell=2$ for the 10~\msol\ CLES models in Fig.~\ref{fig:xprofil}. Dots represent the differences between theoretically predicted  excited modes.}
\label{fig:dnu}
\end{center}
\end{figure}

What is most relevant from an asteroseismic point of view, is the change in the distance between consecutive frequencies of g-modes or modes of mixed p-g character ($\Delta\nu$). In Fig.~\ref{fig:dnu} we  plot these differences for the $\ell=2$ modes of the 10~\msol\ models with an effective temperature \teff$\simeq$22550~K ($X_{\rm C}\simeq 0.3$). The dots indicate differences computed between pairs of excited modes. Therefore the difference of $\Delta \nu$ that we can expect between models with sharp $\nabla_\mu$ (overshooting models for instance) and models with a chemical composition gradient smoothed by the effect of, for instance, a slow rotation ($V_{\rm rot}\sim 50$~km/s) is of the order of 0.4~c/d ($\sim 5\,\mu$Hz), much larger than the precision of present and forthcoming  observations.

\begin{figure}
\begin{center}
\includegraphics[width=8cm]{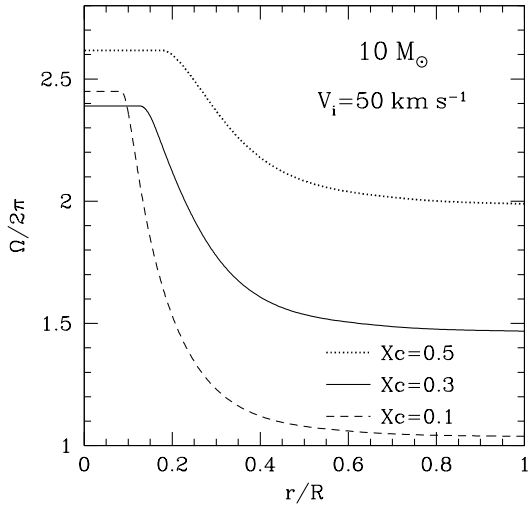}
\caption{Angular velocity profile as a function of the normalized radius for a Geneva 10~\msol\ model at three different evolutionary stages, and with an initial rotational velocity $V_{i}=50$~km~s$^{-1}$.}
\label{fig:rotation}
\end{center}
\end{figure}

\begin{figure}
\begin{center}
\includegraphics[width=8cm]{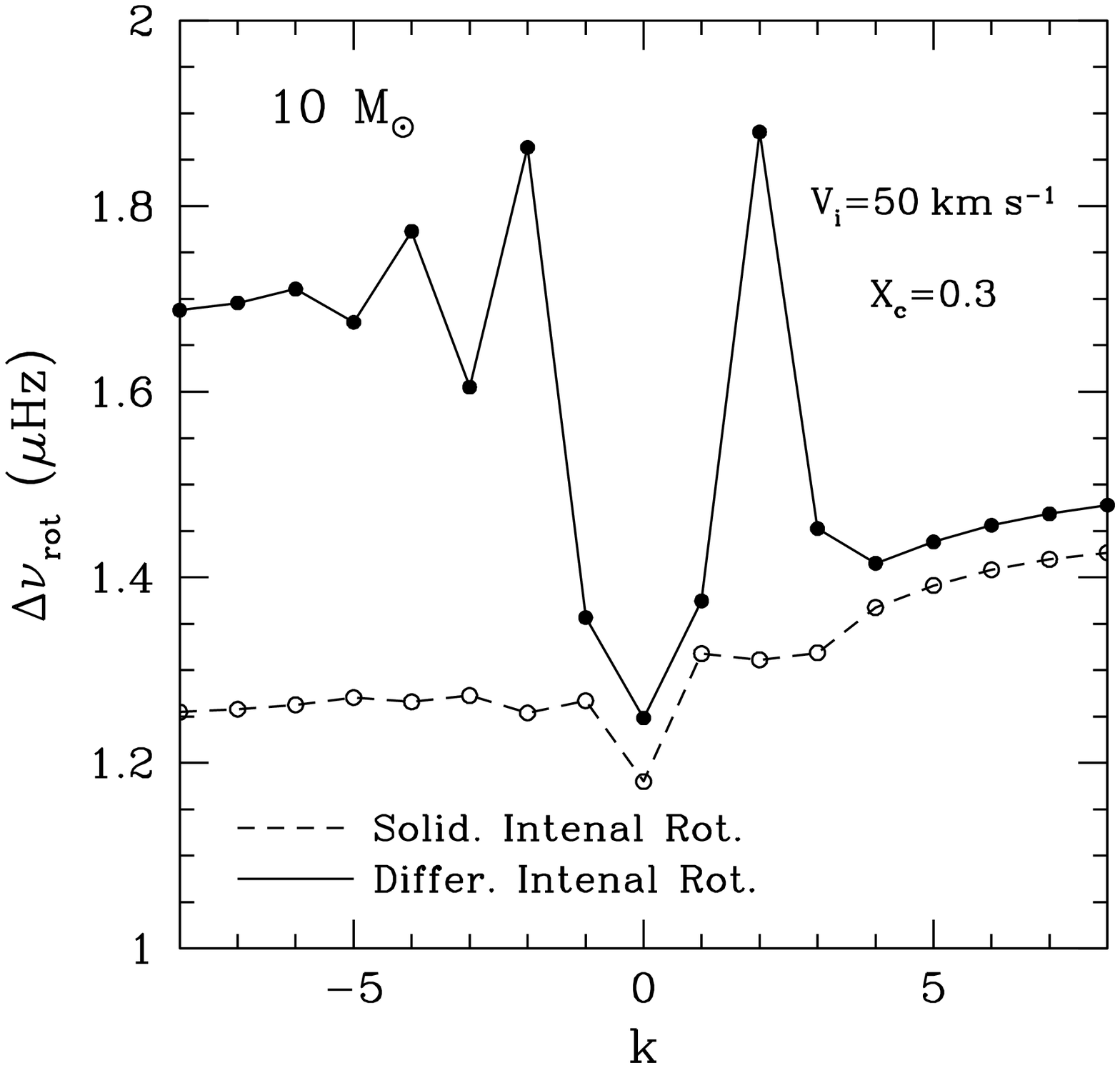}
\caption{First order rotational splitting for $\ell=2$ modes of radial order $k$, for  a 10~\msol\ model at $X_{\rm C}=0.3$ and with a uniform internal rotation rate (dashed line) and with the corresponding  internal rotation profile give in Fig.~\ref{fig:rotation} (solid line).}
\label{fig:dnurot}
\end{center}
\end{figure}

\subsection{Effects of rotation}
\label{sec:rot}
Up to now we have considered the effect of changing  $\nabla_\mu$ by chemical mixing on the oscillation frequencies, regardless of the physical processes at the origin  of that chemical  transport. In the last ten years, the stellar models including  the rotationally induced mixing based on the theory by \cite{Zahn92} have been very successful, particularly for massive stars, in explaining different observational facts indicating chemical mixing in the stellar radiative regions \citep[e.g.][and references therein]{MM05}. It is beyond the scope of this paper to  study of the effects of rotation on the oscillation frequency spectrum. Detailed studies of the first, second and third order corrections of frequencies, as well as the mode coupling due to rotation have been addressed in e.g. \cite{Suarez06}, \cite{jagoda02}, \cite{soufi98} and \cite{DG92}. We intend here to show that if the mechanism responsible for chemical mixing is similar to that described in \cite[][and references therein]{Maeder03} additional effects will appears in the oscillation spectrum. In fact, Geneva models also account  for the transport of angular momentum and predict the rotational profile in the interior of the star as a function of time. On the bases of the high anisotropic turbulence \citep{Zahn92} the profile of rotational velocity generated can be assumed to be shellular 
($\Omega=\Omega(r)$). In that case, the  adiabatic oscillation frequencies corrected by the first order effects are given by:

\begin{equation}
\sigma_{k \ell m}=\sigma_{k \ell 0} + m \int K_{k \ell} \Omega(r)dr,
\end{equation}

\noindent 
where $K_{k \ell}$ are the rotation kernels based on the eigenfunctions of non-rotating model \citep{ostriker67} with eigenfrequencies $\sigma_{k,\ell}$.

The rotational profile for three different evolutionary stages of the Geneva 10~\msol\ models with an initial rotational velocity $V_{i}$=50~km~s$^{-1}$ are shown in Fig.~\ref{fig:rotation}. The gradient of angular velocity between the convective core and the external layers increases with time. The ratio between angular velocity in the convective core and at the surface goes from 1.3 at $X_{\rm C}=0.5$,  1.6 at $X_{\rm C}=0.3$, and 2.5 at $X_{\rm C}=0.1$.

We have estimated the first order rotational splitting for a uniform and a  differential rotation. In the former we assume an angular velocity equal to the surface value at that evolutionary stage, and in the latter we use the rotational profile in Fig.~\ref{fig:rotation}.  The effect of rotation does not significantly depend on the degree $\ell$ (for $\ell <3$) \citep[see][]{Suarez06}. In Fig.~\ref{fig:dnurot} we plot the rotational  splitting for $\ell=2$  and $|m|=1$ modes with solid and differential rotation. Since the treatment of rotation evolution predicts a core rotating more rapidly than the surface layers, the g-modes and g-p mixed modes  in the differential rotational case show a larger splitting than p-modes  or modes (such as the one with $k$=-1) whose eigenfunction has the properties of an acoustic one with high amplitude in the external region. On the other hand, the mode with $k$=2 is mainly concentrated in the central stellar regions. 
The rotational splitting for these modes ranges from 0.5 to 2~$\mu$Hz. 

There are of course also  second order effects that affect the oscillation frequencies of rotating stars. However, for the low-order modes expected in $\beta$~Cep stars, the term $\Omega/\sigma$ in 50 km/s models is of the order of $2\times10^{-2}$.

For the SPB high-order modes, the Geneva 6~\msol\ model with $X_{\rm C}=0.3$ and an initial rotation velocity $V_i=25$~km~s$^{-1}$ provides a stellar structure with a core that rotates  1.8 times faster than the surface. However, given that in this  model the rotationally induced chemical mixing is able to remove the sharpness in the \BV\ frequency profile, no high-order g mode is particularly trapped in a confined region near the core (see Miglio et al.2007, this proceedings) and therefore all the modes  expected to be excited in SPB pulsators show similar rotational splitting. The splitting difference between uniform and differential rotation is of the order of 30\% and corresponds to the different angular velocity between the surface and the region where the SPB modes have their maximum amplitude ($r/R \sim 0.3$). So, the rotational splitting for a typical SPB is $\Delta\nu_{\rm rot} \simgt 1$~$\mu$Hz  in the oscillation frequency domain going from 4 to 12~$\mu$Hz.
Thus, even in the case of differential rotation we do not expect, in the first order correction, a significantly different rotational splittings for different modes.
However, in SPB's frequency domain the ratio $\Omega/\sigma > 0.1$ and therefore second and third order effects of rotation may be important leading to asymmetric splitting of frequency modes.

\section{Conclusions}
\label{sec:conclu}
 By using a parametric modeling of turbulent mixing in the central region of B-type stars we have shown that the effect of varying the chemical gradient on the oscillation frequencies  are very significant  (5~$\mu$Hz  for instance, for moderate mixing $D_{\rm T}=7\times10^4$~cm$^2$~s$^{-1}$). This effect on the frequencies is in fact  larger or of the same order than the rotational splitting for modes with a moderate rotational velocity typical of $\beta$~Cep (50~km~s$^{-1}$). We recall that the precision expected from forthcoming  observations with COROT satellite \citep{Baglin98} is of the order of 0.1~$\mu$H. Therefore, in asteroseismic modeling, in addition to the rotational splitting, the shifts of frequencies produced by a possible turbulent mixing should be taken into consideration.

For high-order g-modes in SPB stars, the frequency differences between consecutive modes due to the trapping in the sharp $\nabla \mu$ regions is of the order of 0.08--0.7~$\mu$Hz \citep{miglio07}. This difference almost dissapears when a slight turbulent mixing ($D_{\rm T}=5000$~cm$^2$~$s^{-1}$), that could  result from  the rotationally  induced mixing in a star that has evolved from an initial rotation velocity of 25~km~s$^{-1}$, acts in the central stellar region.
Rotation has however other significant effects on  SPB oscillation frequencies: a  symmetric splitting $\Delta \nu _{\rm rot} > 1\,\mu$Hz due to  first order corrections, and   second and higher order corrections that can be significant for the low frequencies in SPBs.

The results presented in previous sections for $\beta$~Cep and SPB stars are qualitatively valid also for $\delta$~Scuti and $\gamma$~Dor pulsators respectively. The situation is however  even more complicate for those cases since these less massive stars rotate typically  faster than $\beta$~Cep  and SPB.

\acknowledgements
J.M. and A.M. acknowledge financial support from the European Helio- and Asteroseismology Network HELAS, from the Prodex-ESA Contract Prodex 8 COROT (C90199) and from FNRS. P.E. is thankful to the Swiss National Science Foundation for support.
\newpage

\bibliographystyle{aa}
\bibliography{montalban}
 
\end{document}